# Nonextensive Thermodynamics and Glassy behaviour in Hamiltonian systems

*Andrea Rapisarda*[*] *and Alessandro Pluchino* [§]
*Dipartimento di Fisica e Astronomia and INFN - Università di Catania, 95123 Catania, Italy*
[*] andrea.rapisarda@ct.infn.it   [§] alessandro.pluchino@ct.infn.it

When studying phase transitions and critical phenomena one usually adopts the canonical ensemble and exploits numerical Monte Carlo methods to predict the equilibrium behaviour. However it is also very interesting to follow the microcanonical ensemble and use molecular dynamics to investigate how the system reaches equilibrium. This is particularly true for finite systems and long-range interaction models since, in this case, extensivity and ergodicity are not assured and deviations from standard thermodynamics are usually found. On the other hand recently many data are available for phase transitions in finite systems as for example in the case of nuclear multifragmentation, or atomic clusters, and there is also much interest in studying plasma and self-gravitating systems [1]. Moreover the generalized thermodynamics introduced by Constantino Tsallis [2] to explain the complex dynamics of nonextensive and non-ergodic systems provides further stimuli and a challenging test in the same direction. In this context an instructive and apparently simple model of fully-coupled rotators, the so-called *Hamiltonian Mean Field* (HMF) model, was introduced in 1995 and intensively studied in the last decade [3-11] together with a generalized version with variable interaction range [12]. Such Hamiltonians have revealed a very complex out-of-equilibrium dynamics which can be considered paradigmatic for nonextensive systems [4,13]. We will illustrate in this short paper the interesting anomalous pre-equilibrium dynamics of the HMF model focusing on the novel connections to the generalized nonextensive thermostatistics [10] and the recent links to glassy systems [6,11,14-15].

**The HMF model: a paradigmatic example for long-range N-body classical systems**

The HMF model which we have been studying in detail has an Hamiltonian $H = K + V$, with the kinetic energy $K = \sum_{i=1}^{N} \frac{p_i^2}{2}$ and the potential one $V = \frac{1}{2N}\sum_{i,j=1}^{N}\left[1 - \cos(\vartheta_i - \vartheta_j)\right]$. In the latter $\vartheta_i$ is the orientation angle of the i-*th* spin $\vec{s}_i = (\sin\vartheta_i, \cos\vartheta_i)$ and $p_i$ is the corresponding conjugate coordinate, i.e. the angular momentum or the velocity, since the N spins have unitary mass. All the spins (rotators) interact with each other and in this sense the system is a *mean field model*. The average kinetic term $\bar{K}$ provides information on the temperature of the system, which can be calculated by the relation $T = 2\bar{K}/N$. On the other hand, the potential part V is divided by the total number of spins in order to consider the thermodynamic limit [3]. At equilibrium this Hamiltonian has a second order phase transition: increasing the energy density U=H/N beyond a critical point $U_C = 0.75$, characterized by a critical temperature $T_C = 0.5$, the system then passes from a ferromagnetic (condensed) phase to a disordered (homogeneous) one [3]. In correspondence, the order parameter given by the modulus of the total magnetization, i.e. $M = \left|\sum_{i=1}^{N}\vec{s}_i\right|$, goes from 1 to zero. Using standard procedures one can easily obtain the canonical equilibrium caloric curve,

given by the relation U = T/2 + (1− $M^2$)/2 [3]. On the other hand, by numerically integrating at fixed energy the equations of motion derived from the Hamiltonian, it is also easy to verify that, starting the system close to equilibrium, the simulations reproduce well the theoretical prediction [3].

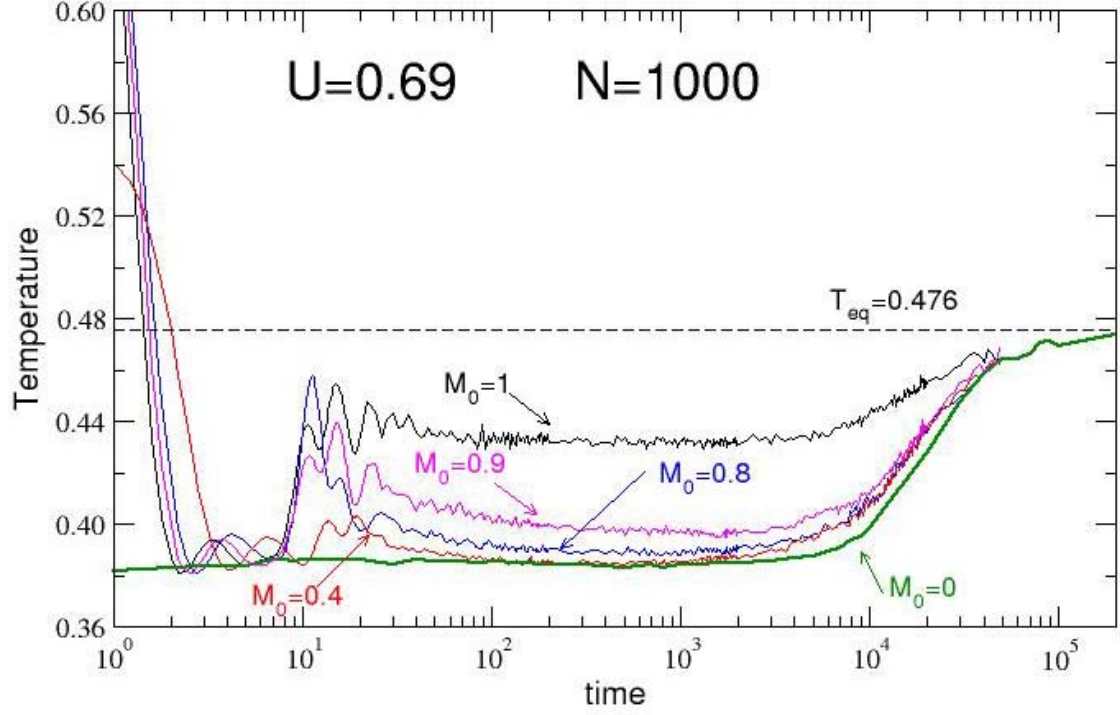

**Fig 1** Time evolution of the temperature extracted from the average kinetic energy for the energy density U=0.69, N=1000 and several initial conditions with different magnetization. After a very quick cooling, the system remains trapped into metastable long-living Quasi-Stationary States (QSS) at a temperature smaller than the equilibrium one. Then, after a lifetime that diverges with the size, the noise induced by the finite number of spins drives the system towards a complete relaxation to the equilibrium value. Although from a macroscopic point of view the various metastable states seem similar, they actually have different microscopic features and correlations which depend in a sensitive way on the initial magnetization.

However, the situation is quite different when the system is started with strong *out-of-equilibrium* initial conditions, as for example giving to the system all the available energy as kinetic one. One way to do this is by considering all the angles $\vartheta_i$ =0, thus obtaining an initial magnetization $M_0$=1, and distributing all the velocities in an uniform interval compatible with the chosen energy density –*water bag* distribution [3]. We have shown that adopting such initial conditions and for an energy density interval below the critical point, i.e. $0.5 \leq U \leq U_C$, the microcanonical dynamics does have difficulties in reaching Boltzmann-Gibbs equilibrium: in fact, after a sudden relaxation from an high temperature state, the system remains trapped in *metastable long-living Quasi-Stationary States* (QSS) whose lifetime diverges with the system size N [4]. Along these metastable states, the so-called '*QSS regime*', the system is characterized by a temperature lower then the equilibrium one, until, for finite sizes, it finally relaxes towards the canonical prediction $T_{eq}$. But, if the infinite size limit is taken before the infinite time limit, the system never relaxes to Boltzmann-Gibbs equilibrium and remains trapped forever in the QSS plateau at the limiting temperature $T_{QSS}$.

In the last years it has been found that the QSS regime is characterized by many dynamical anomalies, such as superdiffusion and Lévy walks, negative specific heat, non-Gaussian velocity distributions, vanishing Lyapunov exponents, hierarchical fractal-like structures in Boltzmann μ-space, slow-decaying correlations, aging and glassy features [3-12].

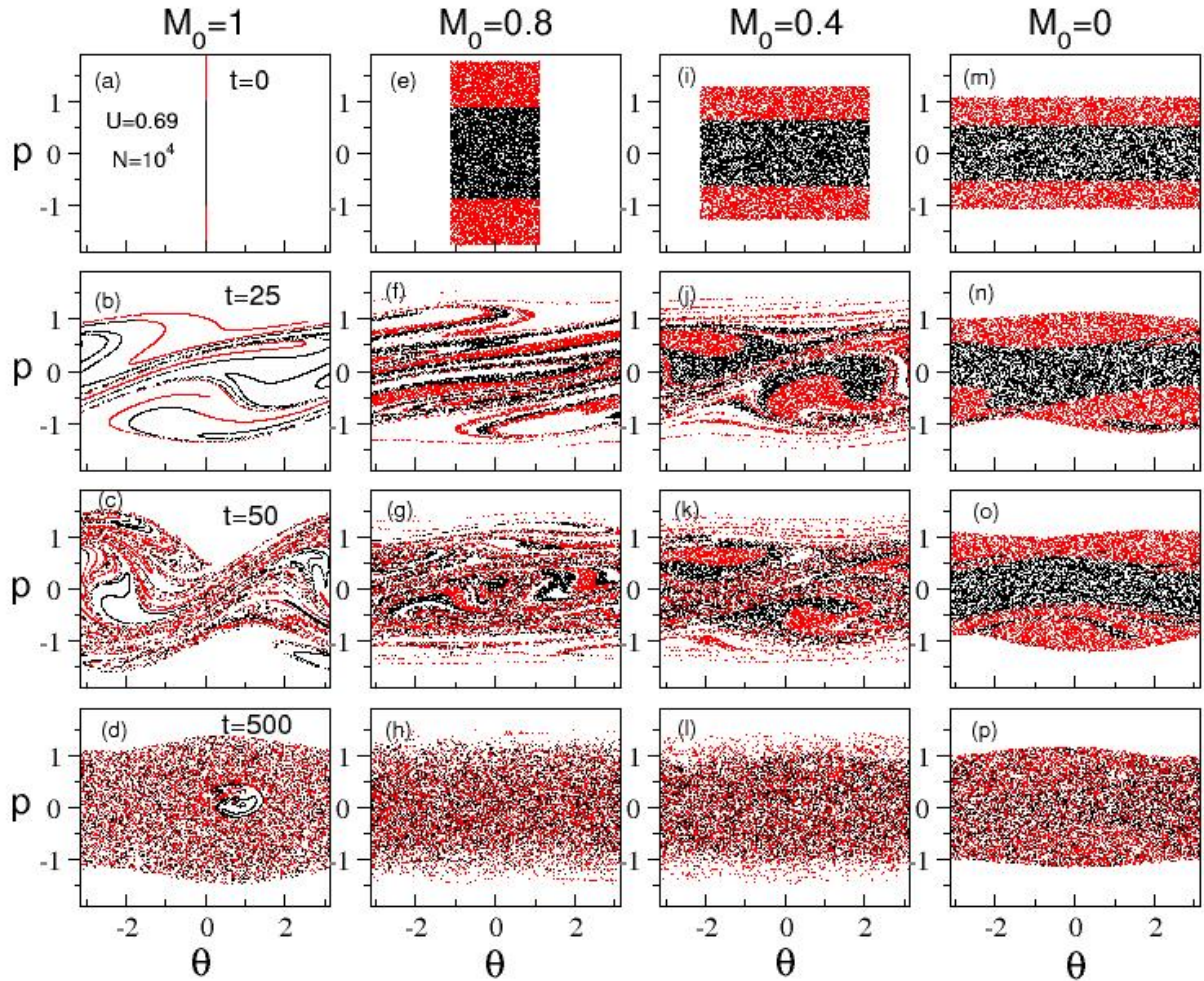

**Fig 2** Initial time evolution of $\mu$-space at four times (t=0,25,50,500) and for different initial magnetizations. The energy density is U=0.69 and the number of spins is N=10000. The initial fast particles are plotted in red to illustrate the mixing properties of the dynamics in the various cases. Fractal-like structures are formed and persist in the QSS regime for $M_0>0$. On the other hand, when the initial magnetization is $M_0=0$ the system remains in a homogeneous configuration. In this case microscopic correlations are almost absent and no structure is evident.

It has been also observed [5] that these anomalies strongly depend on the initial magnetization $M_0$. In *Fig.1* we show the QSS temperature plateaux for U=0.69 (a value for which the anomalies are more evident), N=1000 and for different out-of-equilibrium initial conditions with $0 \leq M_0 \leq 1$. The latter are realized by spreading the initial angles $\vartheta_i$ over wider and wider portions of the unit circle and using uniform distribution for the momenta. One can see from the figure that the system starts from an initial temperature value that rapidly decreases according to the initial magnetization $M_0$, until it reaches the metastable QSS. Only for $M_0=0$, the system already starts at the limiting plateau corresponding, according to the caloric curve, at a temperature $T_{QSS}=0.38$. In all cases, after a long lifetime, the system relaxes to the equilibrium temperature reported as dashed line. Although the macroscopic metastable states are present for all the initial conditions, from a microscopic point of view the system behaves in a very different way. This is nicely illustrated in *Fig.2* where we report, for the energy density U=0.69 and N=10000, the initial time evolution of the $\mu$-space for four different magnetizations. This figure illustrates how structures emerge and persist in the QSS region, but also their dependence on the initial conditions. In ref. [4] it was shown that fractal-like structures characterize the $\mu$-space for $M_0=1$. On the other hand, these features seems to be smoothed by decreasing $M_0$ until, for $M_0=0$, the microscopic configuration of the system remains always homogeneous. For this reason the latter seems to be the only case where an interpretation of metastability in terms of Vlasov equation [7] could likely be applicable. At variance, Tsallis statistics appears to be the best candidate for all the other cases.

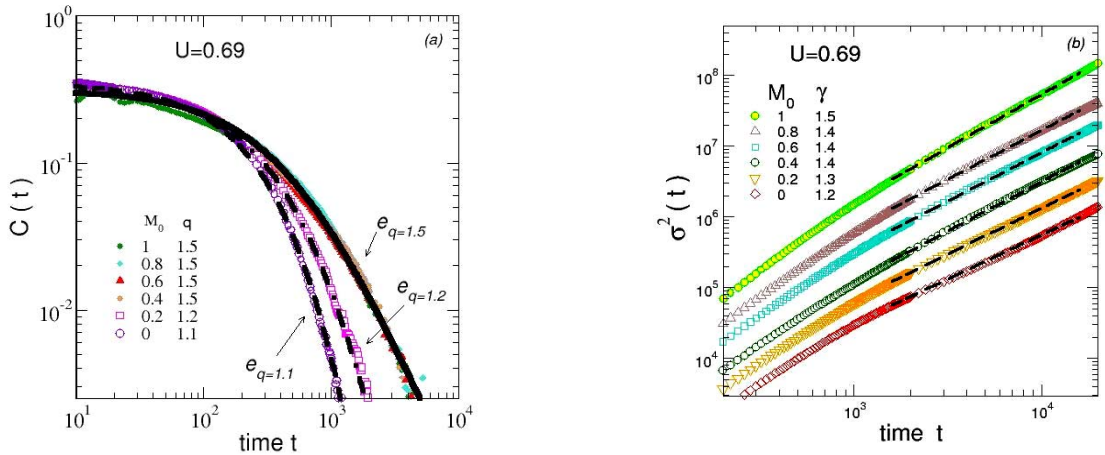

**Fig 3 (a) Time evolution of the autocorrelation functions for U=0.69, N=1000 and different initial conditions are nicely reproduced by q-exponential curves whose entropic index q is also reported. (b) Time evolution of the variance of the angular displacement for U=0.69, N=2000 and different initial conditions. After an initial ballistic motion, the slope indicates a superdiffusive behaviour with an exponent $\gamma$ greater than 1. This exponent is also reported and indicated by dashed straight lines. We have checked that anomalous diffusion does not depend in a sensitive way on the size of the system. For both the plots shown, the numerical simulations are averaged over many realizations.**

**Connections to Tsallis thermostatistics**

In order to explore the characteristic microscopic dynamics originated by the different initial conditions and its connection with Tsallis thermostatistics, one can focus on the *velocity autocorrelation function* $C(t) = \sum_{i=1}^{N} p_i(t)p_i(0)/N$. The latter is plotted in *Fig.3(a)* for U=0.69, N=1000 and several initial magnetizations $M_0$. The initial fast relaxation illustrated in Fig.1 has been truncated to focus only on the properties of the metastable states and an ensemble average over 500 different realizations was performed. We can see immediately that for $M_0 \geq 0.4$ the correlation functions are very similar, while the decay is faster for $M_0=0.2$ and even more for $M_0=0$ [5,11]. These relaxation decays are extremely well fitted by Tsallis' *q-exponential* functions defined as $e_q(z) = [1+(1-q)z]^{1/(1-q)}$, where $z = - t/\tau$, $\tau$ being a characteristic time which indicates the bending of the curve, while *q* is the entropic index [2]. Actually, *nonextensive thermostatistics* introduced by Tsallis has been shown to be particularly adequate to generalize the usual Boltzmann-Gibbs (BG) formalism in describing the out-of-equilibrium dynamics of systems that live in fractal regions of phase space [2]. In this new context, the entropic index q is able to quantify the degree of *nonextensivity* and *non-ergodicity* of the dynamics. For q=1 the standard BG statistics is recovered.
In *Fig.3(a)*, by means of *q*-exponential fits, we illustrate how one can characterize in a quantitative way the dynamical correlations induced by the different initial conditions: in fact we get a value of $q = 1.5$ for $M_0 \geq 0.4$, while $q = 1.2$ and $q = 1.1$ for $M_0=0.2$ and for $M_0 =0$ respectively. Thus for $M_0 \geq 0.4$ correlations exhibit a long range nature and a slow long tailed decay. On the other hand they diminish progressively for initial magnetizations smaller than $M_0=0.4$, to become almost exponential for $M_0=0$. Such a result clearly indicates a different microscopic nature of the QSS in the $M_0=0$ case, which is probably linked to the fact that the latter is a stationary solution of the Vlasov equation [7]. On the other hand, Tsallis' generalized formalism is able to characterize the dynamical anomalies observed not only for $M_0=1$ but also for any finite initial magnetization.

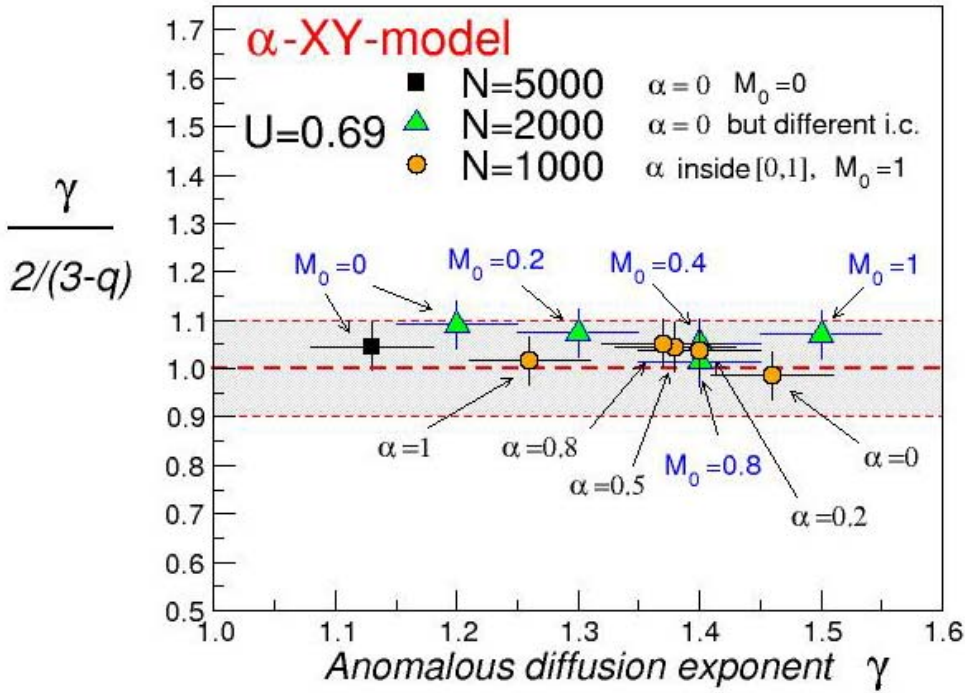

**Fig 4 For different sizes and initial conditions, and for several values of the parameter $\alpha$ which fixes the range of the interaction of a generalized version of the HMF model [12], the figure illustrates the ratio of the anomalous diffusion exponent $\gamma$ divided by 2/(3-q). The entropic index q is extracted from the relaxation of the correlation function (see previous figure). This ratio is always one within the errors of the calculations.**

Actually there are several other results pointing in this direction and in favour of Tsallis generalized statistics [4-6,12-13]. In the following, we want to discuss an interesting conjecture we are currently investigating that gives further support to this interpretation. It concerns a link between the value of the entropic index $q$ which characterize the autocorrelation decay and the exponent of the anomalous diffusion $\gamma$ [10].

In order to observe the diffusion process one can consider the *mean square displacement* of phases $\sigma^2(t)$ defined as $\sigma^2(t) = \langle |\theta_i(t) - \theta_i(0)|^2 \rangle$, where the brackets represent an average over all the N rotators. The mean square displacement typically scales as $\sigma^2(t) \sim t^\gamma$. In general the diffusion is normal when $\gamma = 1$, corresponding to the Einstein's law for Brownian motion, and ballistic (free particles) for $\gamma = 2$. Otherwise, the diffusion is anomalous and in particular one has *superdiffusion* if $\gamma > 1$. In *Fig.2(b)* we plot the mean square displacement versus time for U=0.69, N=2000 and different initial conditions. One can see that, after a ballistic regime proper of the initial fast relaxation, in the QSS regime and afterwards the system clearly shows superdiffusion for $0.4 \leq M_0 \leq 1$ and the exponent $\gamma$ has values in the range 1.4-1.5. On the other hand, in the case $M_0=0$ we get $\gamma=1.2$. Actually we have checked that, by increasing the size of the system, diffusion tends to become normal ($\gamma=1$ for N=10000). Again this case seems to be quite peculiar and microscopically very different from the others studied, where anomalies are much more evident.

We have recently found that superdiffusion observed in the slow QSS regime seems to be linked with the q-exponential decay of the velocity correlations through what we call the *'γ-q conjecture'*, based on a generalized nonlinear Fokker-Planck equation that generates q-exponential space-time distributions [10]. In this framework the entropic index $q$ is related to the parameter $\gamma$ by the relationship $\gamma = 2/(3-q)$. Since in diffusive processes space-time distributions are linked to the respective velocity correlations by the well known Kubo formula, one could investigate the *γ-q* relation considering the entropic index $q$ characterizing the velocity correlation decay in an anomalous diffusion scenario. In *Fig.4* we illustrate the robustness of this conjecture which has been checked numerically by varying not only the initial conditions and the size of the system, but also the range of interaction. These calculations were done by considering the generalized α-XY Hamiltonian, with the parameter α, which modulates the range of the interaction, varying from 0 (HMF model) to 1 [12]. By plotting the ratio $\gamma/[2/(3-q)]$ as a function of $\gamma$ for various values of N, $M_0$ and α at U=0.69, we have checked that the *γ-q conjecture* is confirmed within an error of $\pm 0.1$. This means that knowing the superdiffusion exponent one can predict the entropic index of the velocity correlation decay and viceversa.

The simulations here discussed add an important piece of information to the puzzling scenario of the pre-equilibrium dynamics of the HMF model and its generalizations, which cannot be explained with the standard tools of the BG statistical mechanics. Although these results do not provide a rigorous proof of the applicability of Tsallis generalized statistics, they strongly indicate that this formalism is at the moment the best candidate for explaining the huge number of observed anomalies for a wide class of out-of-equilibrium initial conditions.

**Links to glassy systems**

The importance of the role of initial conditions in generating anomalous dynamics, together with the discovery of aging and dynamical frustration in the QSS regime [6,11], suggests also another non-ergodic description of the HMF dynamics complementary to the Tsallis' one. We are referring to the so-called *weak ergodicity-breaking* scenario for glassy systems [14-15]. The latter occurs when the phase space is a-priori not broken into mutually inaccessible regions in which local equilibrium may be achieved, as in the true ergodicity breaking case, but nevertheless the system can remain trapped for very long times in some regions of the complex energy landscape. It is widely accepted that the energy landscape of a glassy system is extremely rough, with many local minima corresponding to metastable configurations surrounded by rather high energy barriers: one thus expects that these states act as traps which get hold of the system during a certain time. In the

QSS regime of the HMF model, when the system starts from $M_0=1$ initial conditions, such a mechanism is reproduced by the existence of a hierarchical distribution of clusters which compete among each other in trapping more and more particles [11]. Such a phenomenon produces a sort of *dynamical frustration* that recalls the pictorial explanation of aging made by the *cage* model for structural glasses [14] and thus could justify the slow relaxation time observed in the velocity autocorrelation function. Recently, we have extended the analogy between HMF model and glassy systems by introducing a new order parameter, inspired to the Edwards-Anderson (EA) spin-glass order parameter [15], in order to quantify the degree of freezing of the rotators due to the dynamical frustration. We defined the *elementary polarization* as the temporal average, integrated over an opportune time interval $\tau$, of the successive positions of each spin $\langle \vec{s}_i \rangle = \frac{1}{\tau}\int_{t_0}^{t_0+\tau} \vec{s}_i dt$ with $i=1,2,...N$, and $t_0$ being an initial transient time [11].

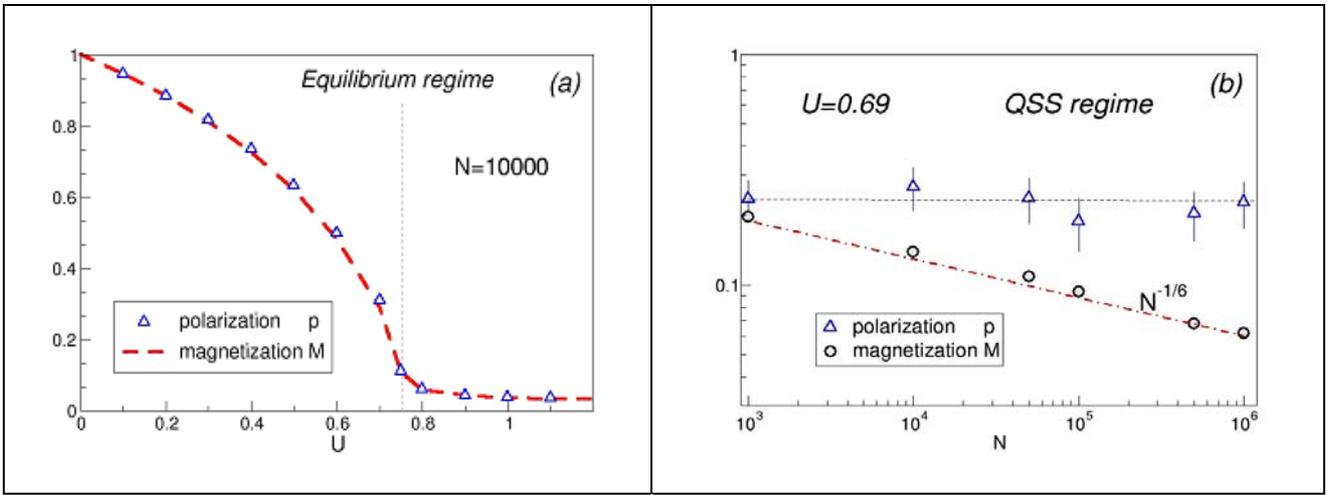

**Fig 5** **(a) The magnetization M and the polarization p are plotted vs the energy density for N=10000 at equilibrium: the two order parameters are identical. (b) The same quantities plotted in (a) are here reported vs the size of the system, but in the metastable QSS regime. In this case, increasing the size of the system, the polarization remains constant around a value $p\sim 0.24$ while the magnetization M goes to zero as $N^{-1/6}$.**

Then we further averaged the module of the elementary polarization over the N spin configurations, to obtain the *polarization p* defined as $p = \frac{1}{N}\sum_{i=1}^{N}|\langle \vec{s}_i \rangle|$. In analogy with the behaviour of the EA order parameter in the Sherrington-Kirkpatrick (SK) model of an infinite range spin-glass [15], we checked that in the equilibrium regime of the HMF model the polarization *p* is equal to the magnetization M. On the other hand, in the out-of-equilibrium QSS regime, which plays here the role of the Spin-Glass (SG) phase in the SK model, the emerging dynamical frustration introduces an effective randomness in the interactions and quenches the relative motion of the spin vectors. Thus *p* has a non null value as in the equilibrium condensed phase, while magnetization, vanishes with the size N of the system and is zero in the thermodynamic limit.
In *Fig.5* we plot the behaviour of *p* and M versus U at equilibrium (a) and in the QSS regime (b) for U=0.69 and the $M_0=1$ initial conditions vs N. One can see that at equilibrium M and *p* assume the same values for both the ferromagnetic and paramagnetic phase. Instead, in the QSS regime magnetization correctly vanishes with $N^{-1/6}$ while polarization remains constant to a value 0.24 ±0.05. This does strongly indicate that we can consider the QSS regime as a sort of glassy phase for the HMF model. Again, it is important to stress the role of the initial conditions in order to have dynamical frustration and glassy behaviour. Actually, we found that glassy features are very sensitive to the initial kinetic explosion that produces the sudden quenching and dynamical frustration. In particular, it has been recently observed [11] that, just reducing $M_0$ from 1 to 0.95 the polarization effect and the hierarchical clusters size distributions become much less evident, until they completely disappear decreasing further $M_0$ . In this sense, the $M_0$=1 initial conditions seems to select a special region of phase space where the system of rotators described by the HMF Hamiltonian behaves as a glassy system. We note in closing that this result gives also a further support to the broken ergodicity interpretation of the QSS regime of Tsallis themostatistics.

**Conclusions**

Summarizing we can surely say that the HMF model and its generalization the α-XY model provide a perfect benchmark for studying complex dynamics in Hamiltonian long-range systems. It is true that several questions remain still open and need to be further studied with more detail in the future. However the actual state of the art favours the application of Tsallis thermostatistics to explain most of the anomalies observed in the QSS regime. The latter seems to have also very interestings links to glassy dynamics.

**Acknowledgments:** We would like to thank V. Latora for his important contribution to this project to which he has been collaborating since the beginning. It is also a pleasure to thank C. Tsallis for many stimulating and encouraging discussions during these exciting past years. Useful discussions with S. Abe, C. Anteneodo, F. Baldovin, F. Bouchet, P-H Chavanis, T. Dauxois, A. Giansanti, I. Giardina, A. Robledo, S. Ruffo and F. Tamarit are also acknowledged.